\documentclass[apjl]{emulateapj}
\usepackage{apjfonts}
\usepackage{color}

\slugcomment{Accepted for publication in ApJL}

\shorttitle{Gamma-rays from W44 Surroundings}
\shortauthors{Y. Uchiyama}

\begin{document}


\title{Fermi-LAT Discovery of GeV Gamma-ray Emission from the Vicinity of SNR W44}


\author{
Yasunobu~Uchiyama\altaffilmark{1,2,3}, 
Stefan~Funk\altaffilmark{1,3}, 
Hideaki~Katagiri\altaffilmark{4}, 
Junichiro~Katsuta\altaffilmark{1}, 
Marianne~Lemoine-Goumard\altaffilmark{5}, 
Hiroyasu~Tajima\altaffilmark{3,6}, 
Takaaki~Tanaka\altaffilmark{3}, 
Diego~F.~Torres\altaffilmark{7,8}
}

\altaffiltext{1}{SLAC National Accelerator Laboratory, 2575 Sand Hill Road M/S 29, Menlo Park, CA 94025, USA.}
\altaffiltext{2}{Panofsky Fellow, uchiyama@slac.stanford.edu}
\altaffiltext{3}{Kavli Institute for Particle Astrophysics and Cosmology, Stanford University, Stanford, CA 94305, USA}
\altaffiltext{4}{College of Science, Ibaraki University, 2-1-1, Bunkyo, Mito 310-8512, Japan}
\altaffiltext{5}{Universit\'e Bordeaux 1, CNRS/IN2p3, Centre d'\'Etudes Nucl\'eaires de Bordeaux Gradignan, 33175 Gradignan, France}
\altaffiltext{6}{Solar-Terrestrial Environment Laboratory, Nagoya University, Nagoya 464-8601, Japan}
\altaffiltext{7}{Institut de Ci\`encies de l'Espai (IEEE-CSIC), Campus UAB, 08193 Barcelona, Spain}
\altaffiltext{8}{Instituci\'o Catalana de Recerca i Estudis Avan\c{c}ats (ICREA), Barcelona, Spain}

\begin{abstract}
We report the detection of GeV $\gamma$-ray emission from 
 the molecular cloud complex that surrounds 
the supernova remnant (SNR)  W44 
using the Large Area Telescope (LAT) onboard \emph{Fermi}. 
While the previously reported $\gamma$-ray emission from SNR W44 
is likely to arise from the dense radio-emitting filaments within 
the remnant, 
the $\gamma$-ray emission  that appears to come 
from the surrounding molecular cloud complex 
can be ascribed to the cosmic rays (CRs) that have escaped from W44.
The non-detection of synchrotron radio emission associated with 
the molecular cloud complex suggests the decay of $\pi^0$ mesons 
produced in hadronic collisions as the $\gamma$-ray emission mechanism.
The total kinetic energy channeled into the escaping CRs is 
estimated to be 
$W_{\rm esc} \sim \mbox{(0.3--3)} \times 10^{50}\ \rm erg$, in broad 
agreement with the conjecture that SNRs are the main sources 
of Galactic CRs. 
\end{abstract}


\section{Introduction}

Diffusive shock acceleration (DSA) operating at expanding shock waves 
\citep[e.g.,][]{MD01} is widely accepted as the mechanism 
to convert the kinetic energy released by supernova explosions 
into the energy of relativistic protons and nuclei (or cosmic rays) 
that obey a power-law type distribution. 
In DSA theory, cosmic rays (CRs) being accelerated at shocks must 
 be scattered by self-generated magnetic turbulence. 
Since the highest-energy CRs upstream in the shock precursor 
are prone to lack self-generated turbulence,  they are expected to escape 
from the shock. 
The character of escaping CRs reflects 
complex interplay between CRs and magnetic turbulence, 
which is yet to be understood. 
DSA theory generally predicts that a substantial fraction of the shock 
energy is carried away by 
escaping CRs, and therefore they
 should be treated as an integral part of the DSA process \citep{EB11}. 

Depending on the amount of nuclear CRs released by a supernova 
remnant (SNR) and the diffusion coefficient in the interstellar medium, 
the molecular clouds illuminated by escaping CRs 
in the vicinity of SNRs are expected to be luminous in $\gamma$-rays 
 due to the enhanced $\pi^0$-decay $\gamma$-rays 
\citep{AA96,Rodri08,AharonianBook,Gabici09}. 
The TeV $\gamma$-ray emission seen in the vicinity of SNR W28 \citep{HESSW28} 
can be regarded as a realization of this scenario.

Recent detections of intense GeV $\gamma$-ray emission from 
SNRs interacting with molecular clouds \citep[e.g.,][]{FermiW51C,FermiW44,FermiIC443,FermiW28,CS10,AGILE_W44} provide 
an interesting opportunity to study how CRs escape from SNRs 
and propagate in the interstellar medium. 
However, it has not been clear whether the $\gamma$-ray emission 
is produced by escaping CRs or by trapped CRs \citep{Uchiyama11}, 
since shocked molecular clouds inside a SNR  also could be the sites of 
efficient $\gamma$-ray production \citep{Bykov00,Uchiyama10}. 

In this Letter,  we report the \emph{Fermi} Large Area Telescope (LAT) 
observations of the region around SNR W44. 
We demonstrate that 
the $\gamma$-ray emission 
distributed on a spatial scale much larger than the remnant 
 is attributable to  a CR halo around SNR W44. 

\section{Observation and Analysis}

\begin{figure*}[htbp] 
\epsscale{0.45}
\plotone{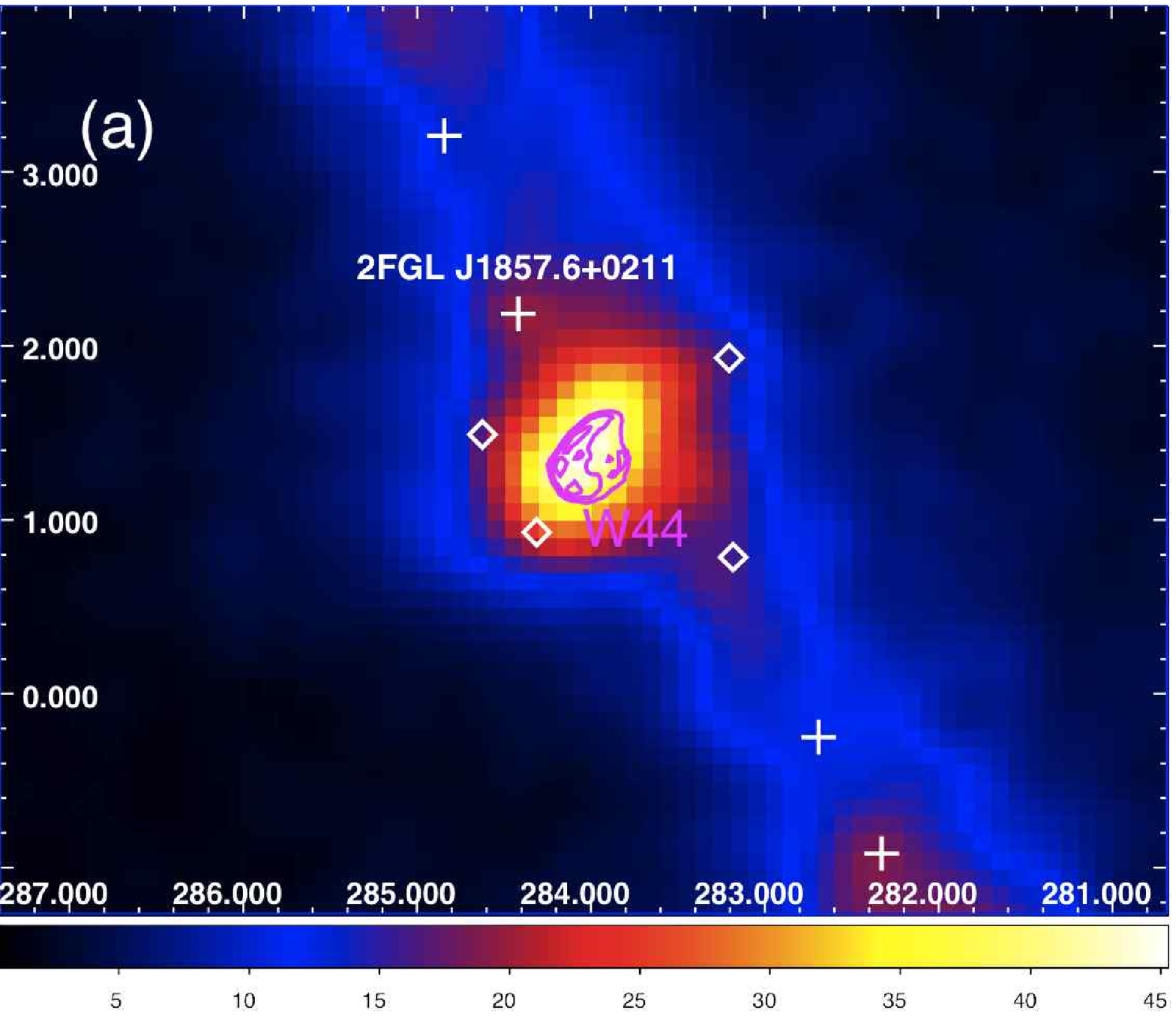}
\plotone{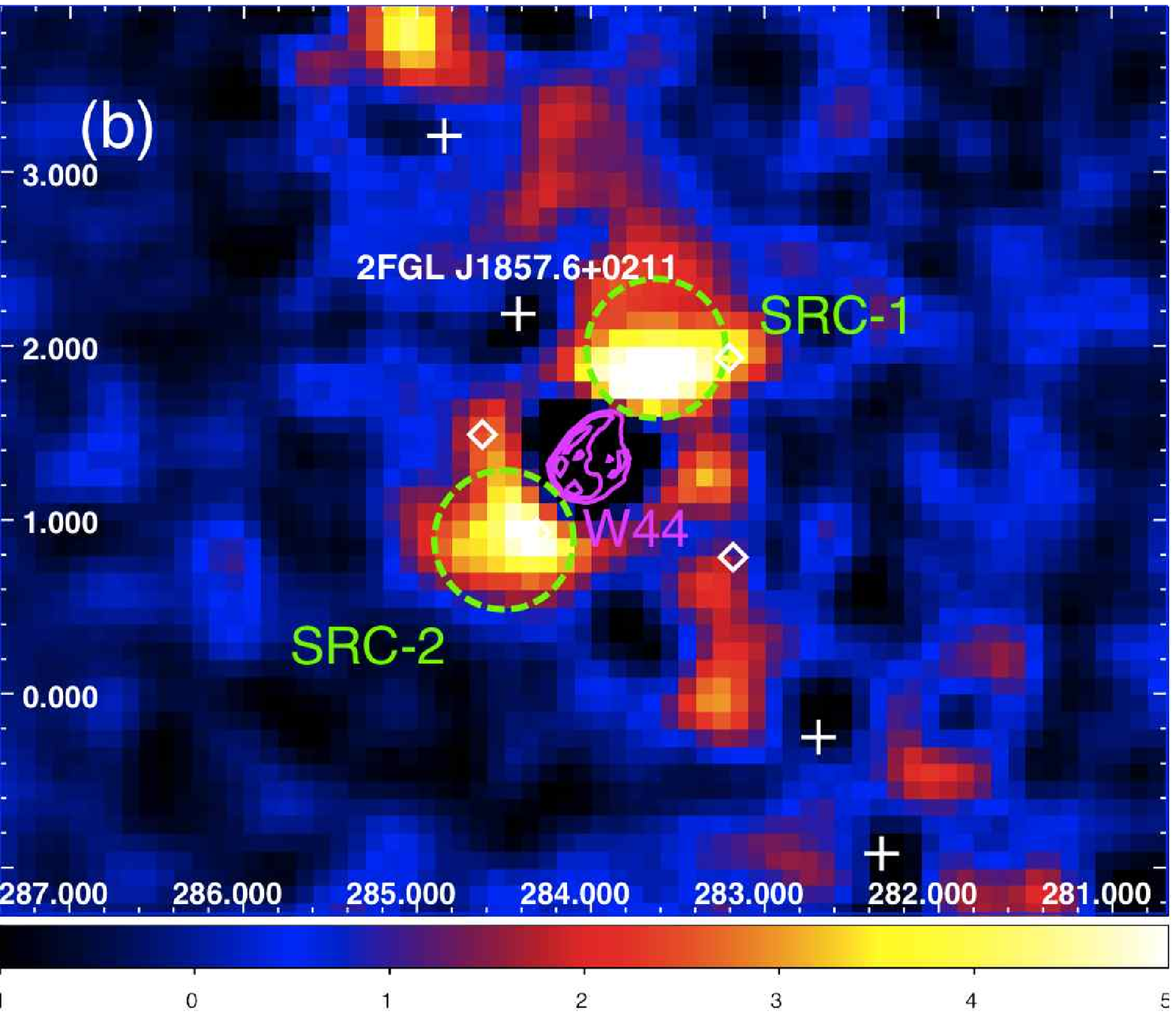}
\caption{
\small 
(a) \emph{Fermi} LAT $\gamma$-ray count map for 2--100 GeV around SNR W44 
in units of counts per pixel ($0\fdg1 \times 0\fdg1 $)
 in celestial coordinates (J2000). 
Gaussian smoothing with a kernel $\sigma = 0\fdg 3$ is applied to 
the count maps. 
Green contours represent a 10 GHz radio map of SNR W44 \citep{Handa87}.
2FGL sources included in the maximum likelihood model are 
shown as crosses, while those removed from the model are 
indicated by diamonds.
(b) The difference between the count map in (a) and the best-fit 
(maximum likelihood) model consisting of 
the Galactic diffuse emission, the isotropic model, 
2FGL sources (crosses), and SNR W44 represented by the radio map. 
Excess $\gamma$-rays in the vicinity of W44 
 are referred to as SRC-1 and SRC-2. 
 \label{fig:cmap}}
\end{figure*}

The LAT onboard \emph{Fermi}
 is a pair-conversion $\gamma$-ray detector that covers 
a very wide range of energy from 20 MeV to $>$300 GeV  \citep{LAT}. 
The LAT tracks the electron and positron  resulting from pair conversion of 
an incident $\gamma$-ray in thin high-$Z$ foils, and measures
the energy deposition due to the subsequent electromagnetic shower that develops in the calorimeter.
The point-spread function (PSF) varies with  photon energy 
and improves  at higher energies. 
The 68\% containment radius is smaller than $0\fdg 5$ above $\sim 2$ GeV. 

We use the $\gamma$-ray data acquired from 2008 August 4 to 2011 September 6.
The \emph{Source} class events are analyzed using 
the instrument response functions (IRFs) {\sf P7SOURCE\_V6}. 
A cut on earth zenith angles greater than $100^{\circ}$ 
is applied to reduce the residual $\gamma$-rays from 
cosmic-ray interactions in the upper atmosphere. 
We analyze photons with energy $\geq  2$ GeV, 
where the PSF is sharp enough to disentangle multiple 
spatial components.
In-depth analysis below 2 GeV will be published elsewhere. 

Spectral and spatial parameter estimation is done by maximizing 
the likelihood of the source model \citep[e.g.,][]{Mattox96} 
using ``binned" 
{\sf gtlike} of  the  \emph{Fermi} Science Tools. 
The region used for the likelihood analysis is 
$10\degr \times 10\degr$,  centered on W44.
The $\gamma$-ray source model includes point sources 
listed in the second  \emph{Fermi} LAT catalog \citep[2FGL sources;][]{2FGL}, 
Galactic interstellar diffuse emission, and an isotropic  component 
(extragalactic and residual particle background). 
The Galactic diffuse emission is modeled using the standard ring-hybrid model, 
{\sf gal\_2yearp7v6\_v0.fits}, 
with its normalization being left free.
We use a tabulated spectrum written in 
{\sf iso\_p7v6source.txt} 
as the isotropic diffuse emission. 
The LAT data, analysis software, 
and diffuse models are made publicly available through the \emph{Fermi} Science 
Support Center\footnote{http://fermi.gsfc.nasa.gov/ssc/}.

Figure~\ref{fig:cmap}(a) shows a 2--100 GeV count map 
 in the vicinity of  SNR W44, where crosses and diamonds 
  indicate the positions of 
 2FGL sources. 
In addition to W44, five 2FGL sources are distributed within 
$1\fdg 5$ from W44. One of them, 
2FGL J1857.6$+$0211, coincides with PSR~B1855$+$02 and also with 
SNR G35.6$-$0.4\footnote{PSR~B1855$+$02 is located near the 
center of G35.6$-$0.4.
At the southern border of G35.6$-$0.4, there is 
a TeV $\gamma$-ray 
source HESS~J1858$+$020 \citep{HESSunID}, toward which 
one or more molecular clouds have been found \citep{PG10}. 
Discussion of HESS~J1858$+$020 can be found in \citet{Torres11}. }
that has recently been re-identified as a SNR \citep{Green09}. 
The other  nearby 2FGL sources  (diamonds) do not 
have clear counterparts in other wavelengths, and they are 
 excluded 
from the  source model to investigate the surroundings of W44. 

We employ a synchrotron radio map of SNR W44 taken from 
\citet{Handa87} to model the  spatial distribution of the $\gamma$-ray
emission from W44, 
given that the synchrotron and $\gamma$-ray emission from W44 
are expected to be co-spatial  (see \S\ref{sec:W44}). 
The $\gamma$-ray spectrum is assumed to obey a power law. 
 
\section{Results}

The likelihood analysis is performed using the source model 
described above. 
For point sources, we use the spectral models adopted for each 
source in the 2FGL catalog analysis. 
Spectral normalizations of point sources located $< 3\degr$ from W44 are allowed to vary in the likelihood fit, while 
the spectral parameters of the other field sources are fixed using 
the 2FGL catalog.
The normalization and photon index of W44 are left free; 
a photon index of $\Gamma_{\rm W44} = 2.94\pm 0.07$ is obtained 
in agreement with our previous work \citep{FermiW44}. 
Figure~\ref{fig:cmap}(b) shows a residual count map, where 
the observed count map in 2--100 GeV is subtracted by the best-fit sky model.
Significant excess $\gamma$-rays are seen in the vicinity of W44; 
the features  are referred to here as SRC-1 and SRC-2. 
The statistical significance 
is found to be $\sim 9\sigma $ for SRC-1 and $\sim 10\sigma $ for SRC-2.

The residual count map depends weakly on the choice of the spatial template 
that describes $\gamma$-rays from W44.  
Our simulations using {\sf gtobssim}  
verified that SRC-1 and SRC-2 are not caused by photons leaking from W44 
due to the PSF of the LAT.
Also we checked 
the robustness of the results by selecting only the front-converted events.

 \begin{figure}[htbp]
  \epsscale{1.2}
   \plotone{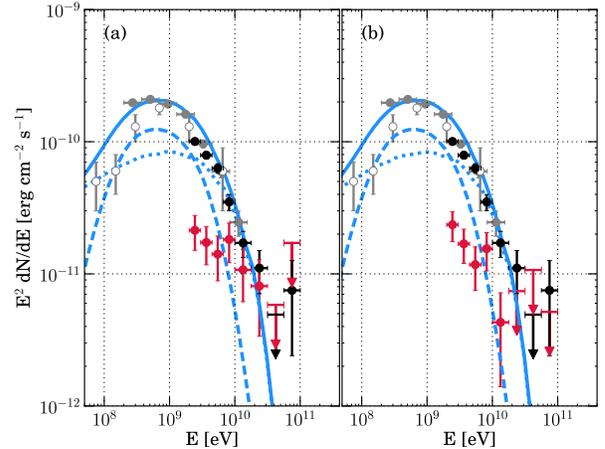}
  \caption{
  \small 
(a) \emph{Fermi} LAT spectrum of SRC-1 (red points) 
along with the LAT spectra of SNR W44 from this work (black points)  
and previous one \citep[][gray points]{FermiW44}. 
The $\gamma$-ray spectrum reported by 
 AGILE is also shown \citep[][open circles]{AGILE_W44}. 
Systematic errors are added in quadrature to the errors of  the SRC-1 spectrum. 
The model curves describe the emission from SNR W44 
(see \S\ref{sec:W44}), consisting of 
 $\pi^0$-decay $\gamma$-rays (dashed curves) and 
relativistic bremsstrahlung (dotted curves). 
(b) Same as (a) but the LAT spectrum of SRC-2 is shown instead of SRC-1. 
  \label{fig:SED}}
 \end{figure}

We perform spectral analysis of SRC-1/2 
by modeling each source as a disk with a $0\fdg 4$ radius (see Fig.~\ref{fig:cmap}).
The resulting $\gamma$-ray spectra are plotted 
 in Figure~\ref{fig:SED} along  with  the spectrum obtained for SNR W44.
Adding the SRC-1/2 disks to the source model does not significantly affect 
the W44 spectrum. 
The power-law photon index is  found to 
be $\Gamma =2.56\pm 0.23_{\rm sta} \pm 0.2_{\rm sys}$ 
and $\Gamma =2.85\pm 0.23_{\rm sta} \pm 0.2_{\rm sys}$ 
for SRC-1 and SRC-2, respectively. 
The systematic errors are evaluated from 
 different choices of the sky models describing W44 and SRC-1/2 
 and from the uncertainties of the effective area. 
 The imperfection of the diffuse emission model 
and its possible impact on the results 
are discussed below.
We tested a smoothly broken power-law and exponentially cutoff power-law 
for SRC-1/2 but found that the spectral fits do not significantly improve. 

\begin{figure}[htbp] 
\epsscale{1}
\plotone{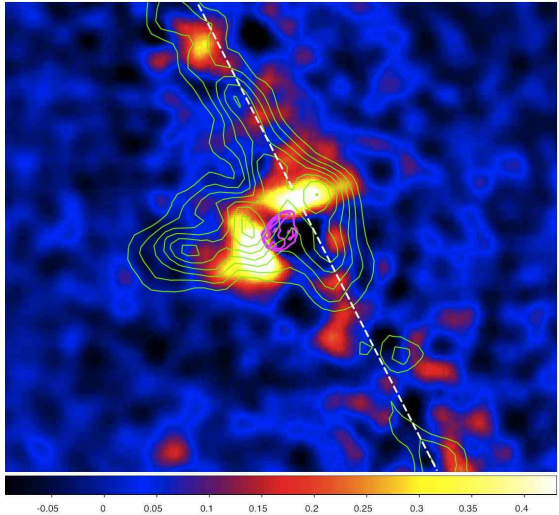}
\caption{
\small 
\emph{Fermi} LAT residual count map highlighting 
the $\gamma$-ray emission from the surroundings of SNR W44. 
Magenta contours present the synchrotron radio map of SNR W44. 
Green contours show 
CO ($J=1\rightarrow 0$) emission integrated over 
velocity from 30 to 65 km s$^{-1}$ with respect to the Local Standard 
of Rest  \citep{Dame01}, tracing 
 the molecular cloud complex that surrounds SNR W44. 
The contours start from 20 K km s$^{-1}$ with an  interval of 10 K km s$^{-1}$. 
Some molecular clouds not associated with W44 are also seen along the Galactic plane (dashed line) in the CO map. 
 \label{fig:CO}}
\end{figure}

\section{Discussion}\label{sec:discuss}

We have discovered GeV $\gamma$-ray sources 
 on the 
periphery of SNR W44. 
It has long been known that a complex of giant molecular clouds (GMCs) 
surrounds SNR W44; the spatial extent is as large as 100 pc 
and the total mass of the complex amounts to $\sim 1\times 10^6 M_\odot$ 
\citep{Dame86,Seta98}.
In Figure \ref{fig:CO}, 
the $\gamma$-ray emission from the surroundings of W44 is 
 compared with a CO ($J=1\rightarrow 0$) map \citep{Dame01} 
integrated over a velocity range of 30 to 65 km s$^{-1}$ appropriate for 
the GMC complex.
The regions of  excess $\gamma$-rays  overlap  with 
the surrounding GMC complex.

The $\gamma$-ray emission in the vicinity of W44 can be ascribed formally 
to possible imperfection of the maps of gas column densities 
used in the model of the Galactic interstellar diffuse emission, 
or to a local enhancement of CR density. 
The former implies that the mass in the $\gamma$-ray-emitting region 
around W44 
is underestimated by a large factor ($\ga 5$), or it requires the presence of 
unknown background 
clouds with a huge mass of $\ga 10^6 M_{\odot}$. 
Therefore an overabundance of CRs in the vicinity of W44 offers a more 
sensible explanation; we present a model in which 
 the GMC complex is illuminated by CRs that 
were produced in SNR W44 and escaped from it. 
We first model the $\gamma$-ray emission 
from W44 itself, and then proceed with 
modeling  the $\gamma$-ray emission from the surroundings.

We adopt the following parameters to describe SNR W44 
 \citep[][and references therein]{Uchiyama10}: 
(i) $d = 2.9\ \rm kpc$ as the distance to W44 based on 
the firm association with the surrounding GMCs \citep[e.g.,][]{Seta98}; 
(ii) $R = 12.5\ \rm pc$ as the radius 
(corresponding to the angular radius of $\theta = 14.8\arcmin$); 
(iii) the kinetic energy released by the supernova 
$E_{\rm SN} = 2\times 10^{51}\ \rm erg$;
(iv) the ejecta mass  $M_{\rm ej} = 2M_{\odot}$; and 
(v)  the remnant age\footnote{The SNR age is comparable to the spin-down age of the 
pulsar B1853$+$01 associated with SNR W44, $\tau_{\rm sd} = 20000$ yr \citep{B1853}.} $t_{\rm age} = 10000\ \rm yr$. 
Assuming evolution in the uniform intercloud medium, 
these parameters imply that 
the intercloud medium has hydrogen density of $n \simeq 2\ \rm cm^{-3}$ and 
the Sedov-Taylor phase started around $t=t_{\rm ST} \simeq 129$ yr when 
the radius was $r_{\rm ST} \simeq 1.9$ pc.

\subsection{Gamma-ray Production Inside SNR W44} \label{sec:W44}

A high-resolution radio continuum map of SNR W44 is dominated by 
filamentary structures  of synchrotron radiation \citep{W44radio}. 
The radio emission is thought to arise from 
radiatively-compressed gas 
behind fast dissociative shocks driven into molecular clouds that 
are engulfed by the blastwave \citep{Reach05}.
Assuming typical magnetic fields of molecular clouds, 
the GeV $\gamma$-ray flux relative to the radio flux is expected to be 
high enough to account for the $\gamma$-ray emission from 
SNR W44, irrespective of the origin of the high-energy 
particles  \citep{Uchiyama10}. 

 \begin{figure}[htbp]
  \epsscale{1.2}
    \plotone{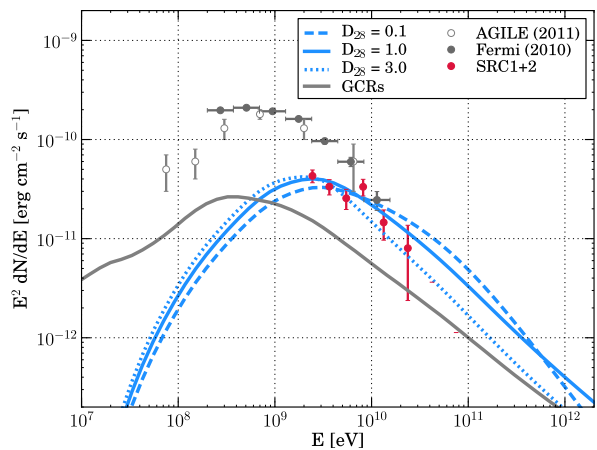}  
\caption{
\small 
 Modeling of the $\gamma$-ray emission from the molecular cloud complex 
 that surrounds W44. Data points are from Fig.~\ref{fig:SED}, but SRC-1 
 and SRC-2 are co-added. 
  The SRC-1+2 spectrum (red points with statistical errors) is attributable to 
the  $\pi^0$-decay $\gamma$-rays from the cloud complex illuminated 
by  the CRs that have escaped from W44 (blue curves). 
Three cases of diffuse coefficient, $D_{28}=0.1, 1, 3$, are shown. 
A gray curve indicates 
the $\gamma$-ray spectrum produced by the sea of GCRs 
in the same CR-illuminated clouds. 
  \label{fig:Model}}
 \end{figure}

The dense radio-emitting filaments are indeed 
the most probable sites of the dominant $\gamma$-ray production in SNR W44, 
given the estimated mass of 
$M_{\rm sh} = 5\times 10^3 M_{\odot}$ \citep{Reach05} 
which is $\sim 9$ times larger than the swept-up intercloud mass. 
Assuming a pre-shock cloud density  
 of $n_0 = 200\ \rm cm^{-3}$ \citep{Reach05} and 
a  pre-shock magnetic field of $B_0 = 30\ \mu$G, 
we can estimate the compressed gas density and magnetic field 
in the  filaments as 
$n_m \simeq 7\times 10^3\ \rm cm^{-3}$ and $B_m \simeq 0.8$ mG following the 
prescription in \citet{Uchiyama10}.

For simplicity we describe the energy distributions of 
CR electrons and protons in the filaments as 
a cut-off power law in momentum:
$n_{e,p}(p) = k_{e,p} p^{-1.74} \exp (-p/p_{\rm c})$, 
where the index is chosen to match the radio spectral index of 
$\alpha \simeq 0.37$ \citep{W44radio}. 
The ratio of radio and $\gamma$-ray fluxes yields $k_e/k_p = 0.05$. 
The spectral break in the LAT spectrum is reproduced by 
$p_{\rm c} = 10\ {\rm GeV}\, c^{-1}$.
As shown in Fig.~\ref{fig:SED}, 
the spectrum below a few GeV is dominated by 
the decays of $\pi^0$-mesons produced in the dense 
filaments\footnote{The AGILE spectral data \citep{AGILE_W44} are not 
taken into account in the model. They will be discussed elsewhere 
in light of a low-energy spectrum measured with \emph{Fermi}.}, while 
the falling part of the LAT spectrum is contributed largely by 
the electron bremsstrahlung. 
The energy density of the CR protons amounts to 
 $u_p \simeq 4\times 10^{2}\ (M_{\rm sh} / 5\times 10^3\, M_\odot )^{-1}\ 
 \rm eV\ cm^{-3}$, which could be explained by 
 reacceleration of Galactic CRs (GCRs) pre-existing in a molecular cloud 
\citep{Uchiyama10}. However, 
we do not specify the dominant source of the $\gamma$-ray-emitting 
particles, since freshly accelerated 
CRs could also enter the filaments 
diffusively from the intercloud medium.
Provided that $u_p$ corresponds to the mean CR density in the shell of the remnant, the total kinetic energy of CRs is 
$W_p \simeq 0.4 \times 10^{50}\ (M_{\rm sh} / 5\times 10^3\, M_\odot )^{-1}\ \rm erg$.

\subsection{Gamma-rays from the Surroundings of W44} 

The CR distributions in the surroundings of SNR W44 
should be determined by (1) how  
CRs are released into the ambient medium, and (2) the diffusion coefficient 
of the interstellar medium, $D_{\rm ISM}(p)$ \citep[e.g.,][]{Gabici09}. 
We consider only CR protons since leptonic emissions are 
unimportant  for an {\it e-p} ratio of $\sim 0.01$.
Also, the non-detection of synchrotron radio emission from the GMC complex 
indicates that electron bremsstrahlung is not the main 
$\gamma$-ray production mechanism. 

Let us assume that 
CRs with a momentum $p$ can escape from the surface of a SNR 
at time $t=t_{\rm esc}(p)$ when the SNR radius becomes $R_{\rm esc}(p)$, 
and that $t_{\rm esc}(p)$ has a power-law form: 
\begin{equation}
t_{\rm esc} (p)  = t_{\rm ST} \left( \frac{p}{p_{\rm max}} \right)^{-1/\chi}, 
\end{equation}
where we adopt $p_{\rm max} = 10^{15}\ {\rm eV}\, c^{-1}$ and 
$\chi = 3$, following \citet{Gabici09} and \citet{Ohira11}. 
This implies that SNR W44 is currently 
releasing CRs with $p = p_0 \simeq 2\ {\rm GeV}\, c^{-1}$.
A simple Sedov-Taylor evolution gives 
$R_{\rm esc} (p)  = r_{\rm ST} ( p/p_{\rm max})^{-2/5\chi}$.

The runaway CR spectrum integrated over SNR expansion 
is expected to be 
of the form $N_{\rm esc}(p)  \propto p^{-2}$ \citep{PZ05}. However, 
depending on the time history of acceleration efficiency and maximum energy, 
it could be different from $p^{-2}$ \citep{Ohira10a,CAB10}. 
We parameterize the total spectrum of CRs injected
 into the interstellar space as 
 $N_{\rm esc}(p)  = k_{\rm esc} p^{-s} \exp (-p/p_{\rm max})$. 

The distribution function of the runaway CRs
at time $t$ at a distance $r$ from the SNR center, $n(p,r,t)$, 
 is described by 
a well-known diffusion equation, which can be solved using 
the method developed by \citet{Atoyan95}. 
We use the following solution of the diffusion equation \citep{Ohira11}:
\begin{equation}
n(p, r, t) = \frac{ N_{\rm esc}(p) }{4\pi^{3/2} R_d R_{\rm esc} r} 
\left[ e^{- (r-R_{\rm esc})^2 / R_d^2 }
-  e^{ - (r+R_{\rm esc})^2/ R_d^2 }   \right], 
\end{equation}
where 
\begin{equation}
R_d (p,t) \equiv 2\sqrt{ D_{\rm ISM}(p) [t-t_{\rm esc}(p)] }. 
\end{equation}
The diffusion coefficient of the interstellar medium is often parameterized as 
\begin{equation}
D_{\rm ISM} (p) = 10^{28}\ D_{28} \ 
\left( \frac{p}{10\ {\rm GeV}\ c^{-1}} \right)^{\delta } \ 
\rm cm^2\ s^{-1}, 
\end{equation}
with constants of $D_{28} \sim 1$ and $\delta \sim 0.6$ based on the 
 GCR propagation model \citep{Ptuskin06}. 
The diffusion coefficient in the close vicinity of a SNR may be 
different from the Galactic average, for example, 
because of Alfv\'en waves generated by CRs themselves \citep{Fujita10}.
We allow $D_{28}$ to vary, but 
fix the index to be $\delta = 0.6$ for simplicity. 
Given that the $\gamma$-ray emission by the escaping CRs is 
visible against SNR W44 in the LAT image, 
we expect the size of the CR halo, 
$R_{\rm CR} (p) = R_d (p, t_{\rm age}) + R_{\rm esc}(p)$, 
 is much larger than the size of the remnant $R= 12.5$ pc. 
Since we find 
$R_{\rm CR} (100\ {\rm  GeV}\, c^{-1}) \simeq 2R$ for $D_{28} = 0.1$, 
we set $D_{28} \geq 0.1$. 
Note however that the constraint depends on the exact choice of $t_{\rm age}$; 
$D_{28} \geq 0.05$ is obtained for $t_{\rm age} = 20,000$ yr.

To calculate the $\gamma$-ray emission from the surroundings of W44, 
we also need to specify the mass distribution. 
The total mass of the cloud complex illuminated by the runaway CRs 
is estimated as $M_{\rm MC} \sim 5\times 10^5\, M_\odot$
by adding up six molecular clouds (Clouds 1--6) in   \citet{Seta98}.
The mass is  assumed to be uniformly distributed within 
the radius of the molecular cloud complex, $R_{\rm MC} = 50$ pc.
Given the spherical symmetry of the model, 
we combine the $\gamma$-ray spectra of SRC-1 and SRC-2. 

Effectively, our model has three adjustable 
parameters, $D_{28}$, $k_{\rm esc}M_{\rm MC}$, 
and $s$. 
As shown in Fig.~\ref{fig:Model},
we determine $k_{\rm esc}M_{\rm MC}$ (i.e., the normalization) and 
$s$ to reproduce the $\gamma$-ray spectrum  
for three values of the diffusion coefficient: $D_{28} = 0.1, 1, 3$. 
For slow diffusion of $D_{28}=0.1$, a steep index of $s=2.6$ is 
required since CR protons with  $p \la 400\ {\rm  GeV}\, c^{-1}$ 
are still within the GMC complex. 
CRs at higher energies are leaking from 
the complex, and correspondingly the $\gamma$-ray spectrum exhibits 
a break at $\sim 50$ GeV. Though the steep injection index required 
for the slow diffusion case is not theoretically expected \citep{CAB10}, 
this problem may be alleviated by adopting a larger value of $\delta$. 
The total amount of CRs released into interstellar space 
amounts to $W_{\rm esc} \simeq 0.3\times 10^{50}\ 
(M_{\rm MC} / 5\times 10^5\, M_\odot )^{-1}\ \rm erg$.
For a nominal value of $D_{28} = 1$, 
we obtain $s=2.0$ and 
 $W_{\rm esc} \simeq  1.1\times 10^{50}\ 
(M_{\rm MC} / 5\times 10^5\, M_\odot )^{-1}\ \rm erg$; 
 the $\gamma$-ray spectrum in the LAT range is steepened 
due to energy-dependent escape from the GMC complex. 
Finally, 
we find $s=2.0$ and 
 $W_{\rm esc} \simeq 2.7\times 10^{50}\ 
(M_{\rm MC} / 5\times 10^5\, M_\odot )^{-1}\ \rm erg$ to reproduce the 
spectrum in the case of  $D_{28} = 3$.

We estimate that the energy channeled into 
the escaping CRs amounts to 
$W_{\rm esc} \sim \mbox{(0.3--3)} \times 10^{50}\ \rm erg$. 
On the other hand, 
the CR content trapped in W44 is estimated roughly as 
$W_p \sim 0.4 \times 10^{50}\  \rm erg$ assuming a uniform CR distribution 
between the SNR shell and radio filaments (\S\ref{sec:W44}). 
A combination of the \emph{Fermi}-LAT data 
to be obtained with longer exposures, and future operation of the 
Cherenkov Telescope Array (CTA) will put 
more stringent constraints on the models through 
 determination of 
$\gamma$-ray spatial distributions as a function of energy. 

\acknowledgments
The {\it Fermi} LAT Collaboration acknowledges support from a number of agencies and institutes for both development and the operation of the LAT as well as scientific data analysis. These include NASA and DOE in the United States, CEA/Irfu and IN2P3/CNRS in France, ASI and INFN in Italy, MEXT, KEK, and JAXA in Japan, and the K.~A.~Wallenberg Foundation, the Swedish Research Council and the National Space Board in Sweden. Additional support from INAF in Italy and CNES in France for science analysis during the operations phase is also gratefully acknowledged.







\begin{thebibliography}{}

\bibitem[Abdo et al.(2009b)]{FermiW51C} Abdo, A. A., et al.\  2009b, \apj, 706, L1 
\bibitem[Abdo et al.(2010a)]{FermiW44} Abdo, A. A., et al.\ 2010a, Science, 327, 1103 
\bibitem[Abdo et al.(2010b)]{FermiIC443} Abdo, A.~A., et al.\ 2010b, \apj, 712, 459 
\bibitem[Abdo et al.(2010c)]{FermiW28} Abdo, A.~A., et al.\ 2010c, \apj, 718, 348 
\bibitem[Aharonian(2004)]{AharonianBook} Aharonian, F.~A. 2004, 
Very high energy cosmic gamma radiation: a crucial window on the extreme Universe, World Scientific Publishing
\bibitem[Aharonian \& Atoyan(1996)]{AA96} Aharonian, F.~A., \& Atoyan, A.~M.\ 1996, A\&A, 309, 917 
\bibitem[Aharonian et al.(2008a)]{HESSunID} Aharonian, F. A., et al.\ 2008a, \aap, 477, 353 
\bibitem[Aharonian et al.(2008b)]{HESSW28} Aharonian, F. A., et al.\ 2008b, A\&A, 481, 401 
\bibitem[Atoyan et al.(1995)]{Atoyan95} Atoyan, A.~M., Aharonian, F.~A., V\"olk, H.~J.\ 1995, \prd, 52, 3265 
 \bibitem[Atwood et al.(2009)]{LAT} Atwood, W.~B., et al.\  2009, \apj, 697, 1071 
 \bibitem[Bykov et al.(2000)]{Bykov00} Bykov, A.~M., Chevalier, 
R.~A., Ellison, D.~C., \& Uvarov, Y.~A.\ 2000, \apj, 538, 203 
\bibitem[Caprioli et al.(2010)]{CAB10} Caprioli, D., Amato, 
E., \& Blasi, P.\ 2010, Astroparticle Physics, 33, 160 
 \bibitem[Castelletti et al.(2007)]{W44radio} Castelletti, G., Dubner, G., Brogan, C., \& Kassim, N.~E.\ 2007, \aap, 471, 537  
 \bibitem[Castro \& Slane(2010)]{CS10} Castro, D., \& Slane, P.\ 2010, \apj, 717, 372 
 \bibitem[Dame et al.(1986)]{Dame86} Dame, T.~M., Elmegreen, 
B.~G., Cohen, R.~S., \& Thaddeus, P.\ 1986, \apj, 305, 892 
 \bibitem[Dame et al.(2001)]{Dame01} Dame, T. M., Hartmann, D., \& Thaddeus, P. 2001, \apj, 547, 792
\bibitem[Ellison \& Bykov(2011)]{EB11} Ellison, D.~C., \& Bykov, A.~M.\ 2011, \apj, 731, 87 
\bibitem[Fujita et al.(2010)]{Fujita10} Fujita, Y., Ohira, Y., 
\& Takahara, F.\ 2010, \apjl, 712, L153 
\bibitem[Gabici et al.(2009)]{Gabici09} Gabici, S., Aharonian, F.~A., \& Casanova, S.\ 2009, \mnras, 396, 1629 
\bibitem[Giuliani et al.(2011)]{AGILE_W44} Giuliani, A., 
Cardillo, M., Tavani, M., et al.\ 2011, \apjl, 742, L30 
\bibitem[Green(2009)]{Green09} Green, D.~A.\ 2009, \mnras, 399, 177 
\bibitem[Handa et al.(1987)]{Handa87} Handa, T., Sofue, Y., 
Nakai, N., Hirabayashi, H., \& Inoue, M.\ 1987, \pasj, 39, 709 
\bibitem[Malkov \& O'C Drury(2001)]{MD01} Malkov, M.~A., \& O'C Drury, L.\ 2001, Reports on Progress in Physics, 64, 429 
\bibitem[Mattox et al.(1996)]{Mattox96} Mattox, J.~R., et al.\ 1996, ApJ, 461, 396 
\bibitem[Nolan et al.(2012)]{2FGL} Nolan, P., et al.\ 2012, \apjs, in press, arXiv:1108.1435 
\bibitem[Ohira et al.(2010)]{Ohira10a} Ohira, Y., Murase, K., 
\& Yamazaki, R.\ 2010, \aap, 513, 17 
\bibitem[Ohira et al.(2011)]{Ohira11} Ohira, Y., Murase, K., 
\& Yamazaki, R.\ 2011, \mnras, 410, 1577 
\bibitem[Paron \& Giacani(2010)]{PG10} Paron, S., \& Giacani, E.\ 2010, \aap, 509, L4 
\bibitem[Ptuskin \& Zirakashvili(2005)]{PZ05} Ptuskin, V. S., \& Zirakashvili, V. N.\
2005, \aap, 429, 755
\bibitem[Ptuskin et al.(2006)]{Ptuskin06} Ptuskin, V.~S., Moskalenko, I.~V., Jones, F.~C., Strong, A.~W., \& Zirakashvili, V.~N.\ 2006, \apj, 642, 902 
\bibitem[Reach et al.(2005)]{Reach05} Reach, W. T., et al. 2005, \apj, 618, 297
\bibitem[Rodriguez Marrero et al.(2008)]{Rodri08} Rodriguez 
Marrero, A.~Y., Torres, D.~F., de Cea del Pozo, E., Reimer, O., 
\& Cillis, A.~N.\ 2008, \apj, 689, 213 
\bibitem[Seta et al.(1998)]{Seta98} Seta, M., Hasegawa, T., 
Dame, T.~M., et al.\ 1998, \apj, 505, 286 
\bibitem[Torres et al.(2011)]{Torres11} Torres, D.~F., Li, H., 
Chen, Y., et al.\ 2011, \mnras, 417, 3072 
\bibitem[Uchiyama et al.(2010)]{Uchiyama10} Uchiyama, Y., 
Blandford, R.~D., Funk, S., Tajima, H., \& Tanaka, T.\ 2010, \apjl, 723, L122 
\bibitem[Uchiyama(2011)]{Uchiyama11} Uchiyama, Y. (on behalf of the Fermi LAT collaboration) 2011, the 25th Texas Symposium on Relativistic Astrophysics, 
arXiv:1104.1197 
\bibitem[Wolszczan et al.(1991)]{B1853} Wolszczan, A., 
Cordes, J.~M., \& Dewey, R.~J.\ 1991, \apjl, 372, L99 
\end{thebibliography}
\end{document}